\documentclass[twocolumn]{aastex62}

\usepackage{amsmath}

\received{September 30, 2019}
\revised{December 20, 2019}
\accepted{December 23, 2019, to appear in ApJS}

\shorttitle{Imaging a pristine CME}
\shortauthors{Rouillard et al. 2020}

\begin{document}

\title{Modeling the Early Evolution of a Slow Coronal Mass Ejection Imaged by the \textit{Parker Solar Probe}}

\author[0000-0003-4039-5767]{Alexis P. Rouillard}
\email{arouillard@irap.omp.eu}

\affiliation{IRAP, Universit\'e Toulouse III - Paul Sabatier,
CNRS, CNES, Toulouse, France}

\author[0000-0002-1814-4673]{Nicolas Poirier}
\affiliation{IRAP, Universit\'e Toulouse III - Paul Sabatier,
CNRS, CNES, Toulouse, France}

\author{Michael Lavarra}
\affiliation{IRAP, Universit\'e Toulouse III - Paul Sabatier,
CNRS, CNES, Toulouse, France}

\author{Antony Bourdelle}
\affiliation{Office national d'études et de recherches aérospatiales, The French Aerospace Lab, Toulouse, France}

\author[0000-0001-8929-4006]{K\'{e}vin Dalmasse}
\affiliation{IRAP, Universit\'e Toulouse III - Paul Sabatier,
CNRS, CNES, Toulouse, France}

\author{Athanasios Kouloumvakos}
\affiliation{IRAP, Universit\'e Toulouse III - Paul Sabatier,
CNRS, CNES, Toulouse, France}

\author[0000-0002-8164-5948]{Angelos Vourlidas}
\affiliation{Johns Hopkins University Applied Physics Laboratory, Laurel, MD 20723, USA}

\author{Valbona Kunkel}
\affiliation{National Weather Service,National Oceanic and Atmospheric Administration, Silver Spring, MD20910, USA}

\author{Phillip Hess}
\affiliation{Naval Research Laboratory, Washington DC, USA}

\author[0000-0001-9027-8249]{Russ A. Howard}
\affiliation{Naval Research Laboratory, Washington DC, USA}

\author[0000-0001-8480-947X]{Guillermo Stenborg}
\affiliation{Naval Research Laboratory, Washington DC, USA}

\author{Nour E. Raouafi}
\affiliation{Johns Hopkins University Applied Physics Laboratory, Laurel, MD 20723, USA}

\begin{abstract}

During its first solar encounter, the \textit{Parker Solar Probe} (\textit{PSP}) acquired unprecedented up-close imaging of a small Coronal Mass Ejection (CME) propagating in the forming slow solar wind. The CME originated as a cavity imaged in extreme ultraviolet that moved very slowly ($<50$ km/s) to the  3-5 solar radii (R$_\odot$) where it then accelerated to supersonic speeds. We present a new model of an erupting Flux Rope (FR) that computes the forces acting on its expansion with a computation of its internal magnetic field in three dimensions. The latter is accomplished by solving the Grad-Shafranov equation inside two-dimensional cross sections of the FR. We use this model to interpret the kinematic evolution and morphology of the CME imaged by \textit{PSP}. We investigate the relative role of toroidal forces, momentum coupling, and buoyancy for different assumptions on the initial properties of the CME. The best agreement between the dynamic evolution of the observed and simulated FR is obtained by modeling the two-phase eruption process as the result of two episodes of poloidal flux injection. Each episode, possibly induced by magnetic reconnection, boosted the toroidal forces accelerating the FR out of the corona. We also find that the drag induced by the accelerating solar wind could account for about half of the acceleration experienced by the FR. We use the model to interpret the presence of a small dark cavity, clearly imaged by \textit{PSP} deep inside the CME, as a low-density region dominated by its strong axial magnetic fields. 
\end{abstract}

\keywords{Slow solar wind (1873), Solar coronal streamers (1486), Solar coronal transients (312)}

\section{Introduction} \label{sec:intro}
\indent The solar atmosphere continually releases coronal material and twisted magnetic fields in the form of Coronal Mass Ejections (CMEs). The three-dimensional (3D) topology and kinematics of CMEs have been studied extensively over the past decade \citep[e.g.][]{Moestl2009,Thernisien2009,Rouillard2010} by exploiting the comprehensive set of remote-sensing and in-situ measurements taken by the \textit{Solar-Terrestrial Relation Observatory} \citep[STEREO;][]{STEREO}. A good understanding of the origin and evolution of these CME properties is a fundamental goal in heliophysics and an absolute necessity to improve space-weather forecasting. The classic picture of a CME observed in white-light images consists of 3-5 parts that evolve dramatically during the eruption and propagation of a CME to 1AU \citep{Vourlidas2012}. They include a shock, sheath, pile-up, cavity, and core. It is thought that most CMEs transport magnetic fields twisted in the form of a magnetic Flux Rope (FR) \citep{Vourlidas2012}. \\

In white-light images, large CME FRs are usually associated with regions of low coronal brightness (or ``cavities'') surrounded by a bright layer of plasma piled up around that dark region. The contour of this ``pile up'' can often be sufficiently bright to be detected by coronagraphs located at different vantage points such as the Large Angle and Spectrometric COronagraph (LASCO) \citep{Brueckner1995} on board the \textit{Solar and Heliospheric Observatory} (\textit{SOHO}) and the Sun Earth Connection Coronal and Heliospheric Investigation (SECCHI) \citep{Howard2008} on board \textit{STEREO}. With assumptions made, the brightness of the boundary of FRs can be used to infer the dimensions and orientation of a CME's magnetic FR \citep{Chen2000ApJ,Thernisien2009}. The smallest transients, such as streamer blobs, can exhibit brightness features reminiscent of FRs and loops, but the cavity is usually not discernible \citep{Rouillard2011}.\\

The continuous tracking of large CMEs from the Sun to spacecraft has provided critical information on how magnetic FRs expand/contract \citep{Moestl2009,Rouillard2010,Rouillard2011,Wood2012}, rotate \citep{Vourlidas2011ApJ,Isavnin2014SoPh,Kay2015ApJ}, and deflect in 3D from the Sun to 1AU \citep{Kay2016ApJ}. For the fast CMEs, the contour of the shock-sheath region that surrounds the FR can also be used to infer the 3D topology of the shock from the corona to the interplanetary medium \citep{Wood2011,Kwon2014,Rouillard2016,Kwon2017,Kouloumvakos2019}. White-light imagery, and heliospheric imagery in particular, thus provides crucial information on the global 3D substructures of the CME. Unfortunately, total brightness images cannot be used to measure the properties of the magnetic field transported by CMEs. However, the distribution of that magnetic field, and of the associated currents inside and around the FR, influence the internal structure and kinematic properties of CMEs that we seek to analyze here. \\

The multipoint \textit{STEREO} mission has definitely validated the croissant-shaped structure of magnetic FRs \citep{Thernisien2009} for a subset at least of CMEs, with an occasional good correspondence found between FR orientations inferred in simultaneous in-situ measurements and white-light imaging \citep{Moestl2009,Rouillard2010,Wood2010}. This important step has fundamental implications for our understanding of the dynamic evolution of this subset of CMEs. However, the difficulties in the more comprehensive analysis by \citet{Wood2017} either challenge the idea that all FRs have a croissant-shaped structure, or alternatively, challenge our current methodology to infer the 3D topology of magnetic fields from either imagery or in-situ data. \citet{Wood2017} note, for instance, that a relaxation of the restrictive force-free field assumption usually employed to reconstruct FRs with in-situ data could lead to significant improvements in our interpretation of the FR properties inferred from in-situ data.\\

A new generation of FR fitting models includes non-force-free assumptions, as well as significant deformation of the internal structure as the FR propagates in the interplanetary medium \citep{Isavnin2016,Nieves2018}. The present paper is the first of a series that seeks to address these points directly and investigate, both observationally and theoretically, the physics that is potentially missing in semi-analytical models and perhaps overseen in the more complete 3D MHD models. 

\begin{figure*}[ht]
\centering
\includegraphics[scale=0.75]{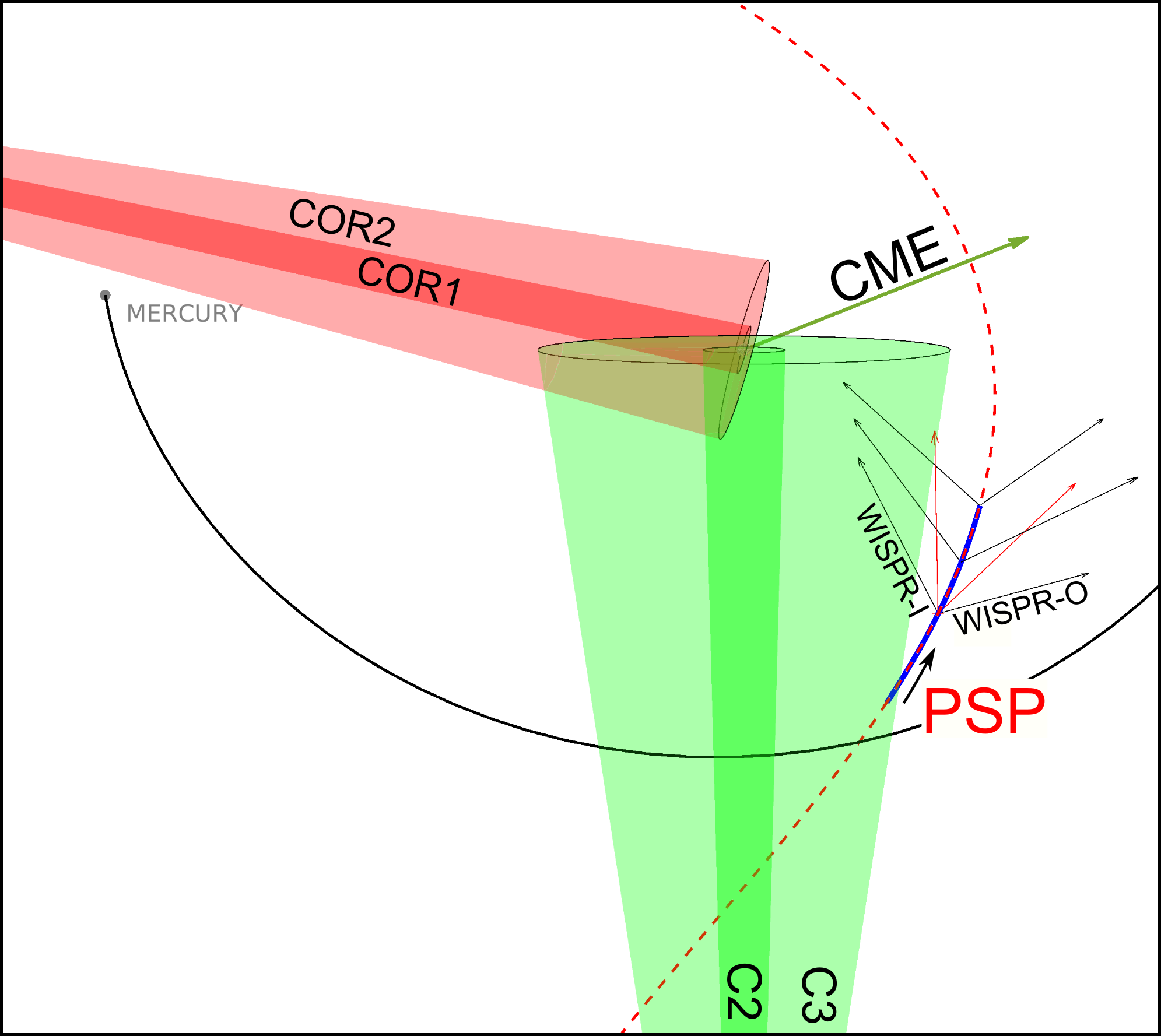}.
\caption{\label{fig:Intro_Fig} A view of the ecliptic plane from solar north, showing the positions of \textit{PSP} on 2018 November 1, 2 and 3. The fields of view of the \textit{SOHO} LASCO C2/C3 and \textit{STEREO}-A COR-1/2 are shown as green and red shaded areas. The thick blue line traces the orbit of \textit{PSP} during the interval of CME observation. The extent of the combined WISPR-I and WISPR-O fields of view are shown for the three dates as black arrows, while the pointing of the center of each camera is shown as a red arrow. The approximate direction of propagation of the CME is shown as a green arrow and corresponds to a longitude at $\sim$115$^\circ$ in  HEEQ coordinates.}
\end{figure*}

CMEs exhibit a broad range of sizes and speeds, with the fastest CMEs accelerating to thousands of kilometers per second in a matter of minutes \citep{Webb2012LRSP}. Fast CMEs typically experience different stages of acceleration, including a gradual-rise stage lasting a couple of hours, followed by a main acceleration stage lasting tens of minutes \citep[e.g.][]{Zhang2001}. The enhancement of the FR's electric current, the increase of the FR twist, and mass losses have been proposed as different but coupled phenomena that can contribute to the initial slow motion of the CME FR \citep{Vrsnak2019}. The latter greater acceleration has been related to a critical height where the FR loses equilibrium \citep[e.g.][]{Chen1989ApJ,Vrsnak1990,Demoulin2010}. \\

A subset of the slowest CMEs can also move very slowly to the outer corona, where they undergo a stronger acceleration to a few hundreds of kilometers per seconds  \citep{Webb2012LRSP}. This second acceleration occurs sometimes tens of hours after their first appearance in the low corona. This paper exploits an eruptive FR model to study the origin of such a long eruption process that was observed during a slow CME imaged by the \textit{Parker Solar Probe} \citep[PSP;][]{Hess2019}. \\

The paper begins with a brief summary of the study carried out by \cite{Hess2019}. We complement this study by carrying out a geometrical fit to estimate the 3D trajectory, kinematic evolution, and expansion rate of the CME. We then describe and exploit a model that computes the forces acting on this FR as it accelerates and expands in the corona. We set the challenge of modeling both the kinematic properties and expansion rates, including the cross-sectional area, of the CME to investigate the possible mechanisms responsible for the two-stage eruption process. We study a number of processes that can influence the emergence of the CME including the torus instability, gravitational buoyancy, and the drag force induced by the background solar wind.\\

\section{Coronal Imaging of the CME} \label{sec:WLobsCMEs}

Figure \ref{fig:Intro_Fig} presents the orbital positions of \textit{PSP} between 2018 November 1 and 3, when the two Wide Imager for Solar Probe \citep[WISPR;][]{Vourlidas2016} instruments were imaging the CME.  The two WISPR telescopes are mounted on the ram side of the spacecraft, and their combined field of view is shown in Figure \ref{fig:Intro_Fig}a as the darker blue area. The combined views cover a range of elongation angles (angular distance from Sun center) from 13.5$^\circ$ to 108$^\circ$ with a spatial resolution of 6.4 arcmin (the images were 2x2 binned). The inner telescope (WISPR-I) extends in elongation angles from 13.5$^\circ$ to 53$^\circ$, and the outer telescope (WISPR-O) extends from 50$^\circ$ to 108$^\circ$ \citep[see][]{Vourlidas2016}. During \textit{PSP}'s first solar encounter, the WISPR instruments obtained full-field and high-cadence images of the corona between 2018 October 31 and November 10 \citep{Howard2019Nat}. At the time, the spacecraft was approaching its first perihelion and WISPR was imaging the solar wind off the west limb of the Sun.  \\

\begin{figure*}[ht]
\centering
\includegraphics[scale=0.5]{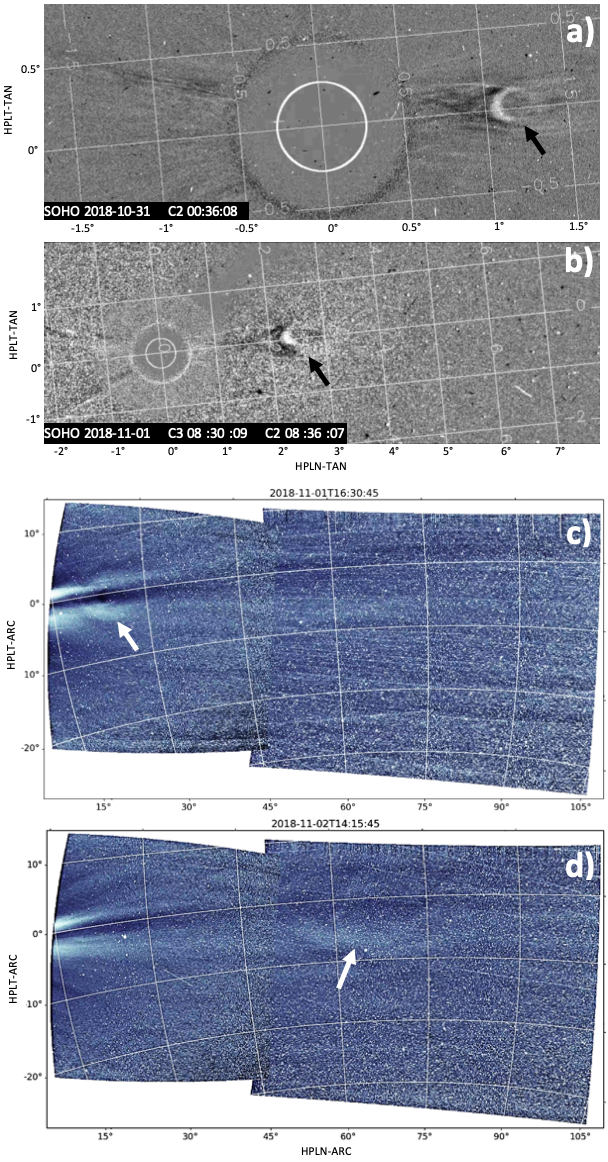}
\caption{Panel (a) shows a running-difference LASCO C2 image of the CME on 2018 October 31 00:36~UTC. Panel (b) displays combined running-difference LASCO C2/C3 images on 2018 November 1 $\sim$08:30~UTC. Panel (c) and (d) show combined Level-3 WISPR I/O images on 2018 November 1 16:30 (2018 November 2 14:15)~UTC. The bright outer boundary of the CME is annotated by black and white arrows. An animation of panels (a) and (b) is available at this \href{https://nuage.irap.omp.eu/index.php/s/8ynNotNVWU6PrtN}{link}. An animation of panels (c) and (d) is available at this \href{https://nuage.irap.omp.eu/index.php/s/dflQ94DKuMbiaTv}{link}. \label{fig:WL_Obs}}
\end{figure*}

Figure \ref{fig:WL_Obs} presents a sequence of running-difference LASCO-C2 (a) and C3 (b) images, as well as combined background-subtracted WISPR-I/O images showing the CME propagation. The technique used to produce images of the K corona from the raw WISPR images is discussed in detail in \cite{Hess2019}. \cite{Hess2019} presented observational evidence for a CME eruption that occurred in two stages. The CME was first observed around 21:00~UT on 2018 October 30 by the Atmospheric Imaging Assembly \citep[AIA:][]{Lemen2012} on board the \textit{Solar Dynamics Observatory} (\textit{SDO}) as the outward motion of a cavity with speeds below 60 km/s. The CME maintained this slow speed up to at least four solar radii (R$_\odot$). This corresponds to about midway inside the LASCO C2 field of view, and a corresponding running-difference image is shown in Figure \ref{fig:WL_Obs}a. A strong acceleration occurred between 4 and 5 R$_\odot$, leading to speeds above 270 km/s as the CME entered the LASCO C3 field of view (Figure \ref{fig:WL_Obs}b). It is not obvious what eruption process would result in an extended slow propagation of the CME in the low corona followed by an acceleration beyond 4 R$_\odot$. We analyze this eruption by combining coronal observations of the CME with a simple model of erupting FRs. \\


The analysis of LASCO C3 (Figure \ref{fig:WL_Obs}b) and \textit{PSP} WISPR-I (Figure \ref{fig:WL_Obs}c) images show the presence of a dark cavity at the center of the CME \citep{Hess2019}. This circular feature is much smaller than cavities observed in images of typical three-part structure CMEs \citep{Vourlidas2012}. We use our model for the internal magnetic field of the FR to investigate the nature and origin of this low-density cavity. The CME morphology also changes rapidly as it progressed in WISPR-O (Figure \ref{fig:WL_Obs}d). The cavity disappears rather abruptly  between 7 and 10 UT on 2018 November 2 as the CME crosses the WISPR-O field of view (FOV) \citep{Hess2019}. This corresponds to a time when the lines of sight from \textit{PSP} can no longer be aligned with the central axis of the CME (toroidal axis), and therefore the detector must have integrated light scattered by plasma located over the entire CME boundary. \\

\citet{Hess2019} show that the FOVs of WISPR-I and LASCO-C3 overlapped in a region of the corona situated off the west limb of the Sun as viewed from Earth. Because of the similarity of the observed features in the region common to both FOVs, they infer that the Thomson spheres of each instrument also overlap, and the two cameras were therefore imaging similar sections of the CME structure at the same time. In both cameras, the CME exhibits a clear outer boundary, especially toward the back of the event where a transition from the bright CME to the corona is clear. In rare cases where such CMEs have been imaged all the way to spacecraft taking in-situ measurements, these bright boundaries were measured as peaks in plasma density immediately adjacent but outside the magnetically dominated regions interpreted as the FRs. In the standard picture of magnetic FRs described in terms of poloidal and toroidal magnetic field components, the bright rim of high plasma density is immediately adjacent to the strong poloidal magnetic fields that maintain the cohesion of the FR.\\

\vspace{1cm}

\section{The 3D geometry of the FR} \label{sec:WLRecCMEs}

\begin{figure}[h!]
\centering
\includegraphics[scale=0.47]{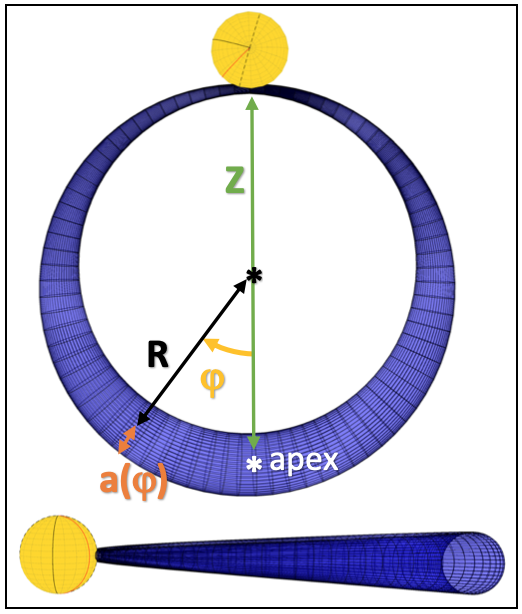}
\caption{ \label{fig:shape} The FR geometry assumed in this study viewed from solar north (top) and from the side (bottom). The different dimensions of the toroidal structure used in equation \ref{eq:aform} are also labeled.}
\end{figure}

Figure~\ref{fig:shape} presents the 3D shape of the CME boundary assumed in this paper to model the CME observed by the LASCO and WISPR instruments. The FR is a bent toroid with a constant major radius, $R$, but a varying minor radius, $a$ with azimuthal angle ($\varphi$). The legs of the FR remain attached to the Sun and have a much smaller cross section at the Sun than the apex of the FR. \\

A constant $R$ means that the FR has a circular symmetry (Figure~\ref{fig:shape}). This `circular current channel' will be considered in section \ref{sec:forces} to calculate the forces acting on such an FR when it erupts in the solar corona. Past studies have found evidence that FRs with noncircular current channels can also successfully fit the aspect of CMEs in coronagraph images taken from different vantage points \citep{Thernisien2009}. In addition, the forces acting on elliptically shaped current channels have also been quantified for ideal cases \citep{Kunkel2012PhDT}. We defer the analysis of these more complex geometries to a future study. \\

In addition to the circularity of the current channel, past studies also assumed that the toroid's minor radius increased either exponentially or linearly with azimuthal angle ($\varphi$) from the footpoints to the apex of the FR. This simplifies the calculation of the inductance of the system, an important step to calculate the forces acting on the FR \citep[see][]{Chen1989ApJ,Chen1996JGR}. These past formulations for the minor radius were justified in the 1D calculation of an FR force balance but cannot be used to produce a 3D representation of the FR. These variations in $a(\varphi)$ lead to discontinuities in the  magnetic flux surfaces of the FR near its apex and prevent a 3D mapping of the internal magnetic field lines. \\

To obtain a more adequate 3D topology, the present study assumes a bell curve for the variation of the minor radius $a(\varphi)$ with azimuthal angle ($\varphi$) measured between the footpoint and the apex: 

\begin{equation}
\label{eq:aform}
a(\varphi)=a_a~exp\Bigg[{-\Bigg(\frac{\varphi}{\varphi_f}\Bigg)^2ln\Bigg(\frac{a_a}{a_f}\Bigg)}\Bigg]
\end{equation}

where $a_f$ and $a_a$ are the minor radii at the footpoint and apex, and $\varphi_f$ is azimuthal angle at the footpoint of the FR. We have retained here a notation similar to that of \cite{Chen1996JGR}, to ease comparison of the different assumed geometries. The minor radius of the FR varies slowly near the apex, with only a 10\% variation of the minor radius along a quarter of the torus centered at the apex. As we shall see, this slowly varying minor radius near the apex of the torus is more consistent with the assumptions made to analytically derive the Lorentz forces acting on the system \citep{Shafranov1966}. \\

We assume this same FR shape to reconstruct the CME evolution in 3D in the next section \ref{sec:Manual_tracking}, and to compute the forces acting on the CME during its eruption process in sections \ref{sec:forces} and \ref{sec:results_FRmodel}.


\section{The 3D reconstruction of the flux rope}
\label{sec:Manual_tracking}
A derivation of the kinematic properties of the CME was performed in \cite{Hess2019} from LASCO C2 to WISPR-I images. They assumed that the CME was propagating in the plane of the sky and measured the extent of the FR cavity and outer boundary, assuming an elliptical FR cross section. In this study, we follow a different approach by carrying out a 3D reconstruction based on the circular FR shown in Figure \ref{eq:aform}. This technique improves on the work of \cite{Hess2019} by correcting for projection effects, to some extent.  \\

\begin{figure*}[ht]
\centering
\includegraphics[scale=0.3]{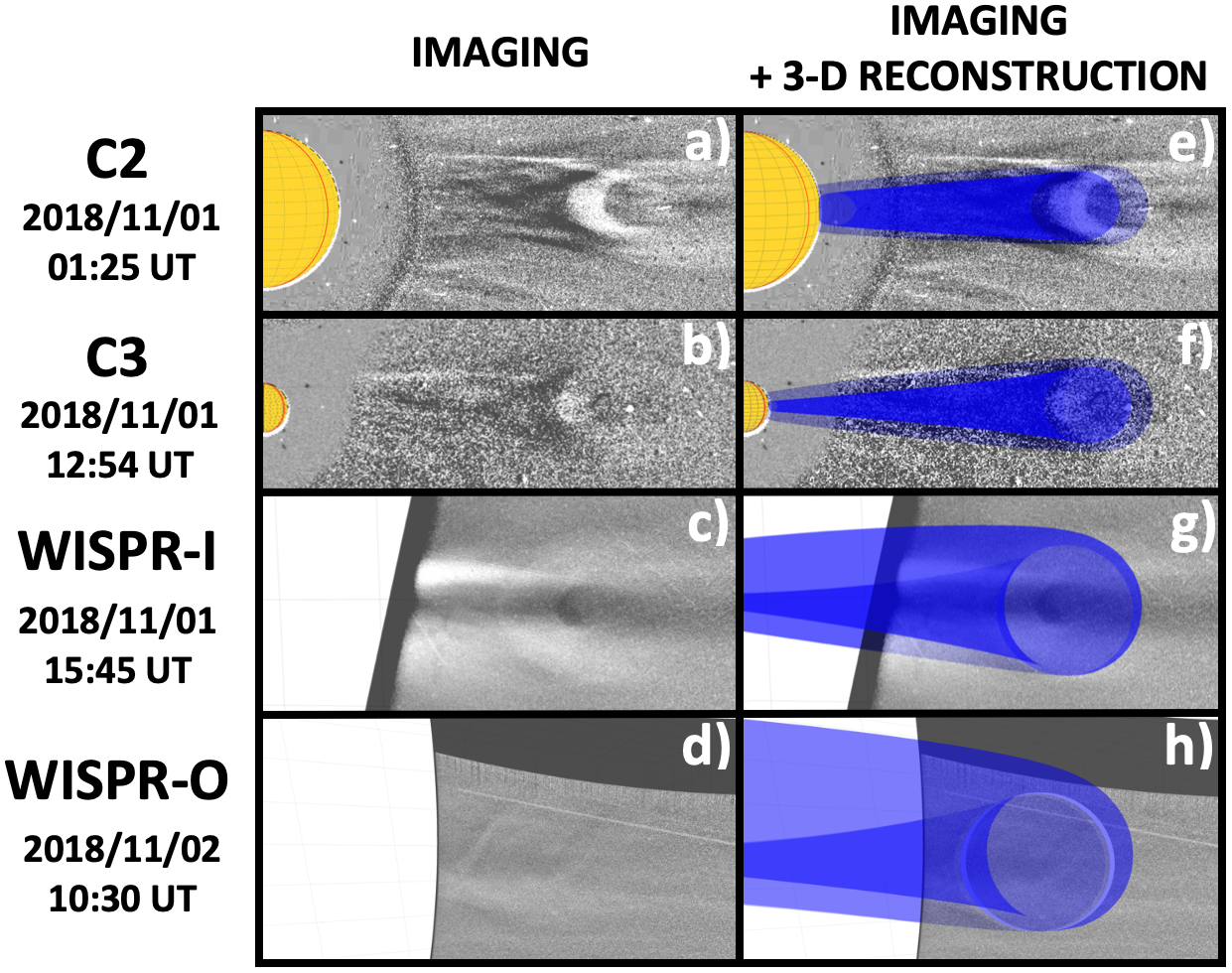}
\caption{Top four panels show the running-difference LASCO C2 image (a and e) and LASCO C3 image (b and f) of the CME. Bottom four panels display WISPR-I (c and g) and WISPR-O (d and h) Level-3 images for which an F-corona removal has been applied \citep[see][]{Stenborg2018ApJ,Hess2019}. On panels (e), (f), (g) and (h), we superimpose the 3D FR fitting (Figure~\ref{fig:shape}) that we use to perform the 3D reconstruction of the CME. \label{fig:Fitting_Obs} }
\end{figure*}

The 3D reconstruction of the FR proceeds in a similar way to the technique of \citet{Thernisien2009}, but assumes the geometry presented in section \ref{sec:WLRecCMEs}. Each image is mapped onto the helioprojective sphere and the FR outline is superimposed on the image by folding in the properties and position of each instrument. The modeled FR can take any desired orientation in 3D until a good visual fit is obtained with the observed CME characteristics. The scene is continually regenerated as the viewing angles of the instruments change along the different spacecraft's orbits. This is essential for \textit{PSP}, which moves very significantly along its orbit during the course of the CME propagation to WISPR-O.
\\
As discussed by \cite{Hess2019}, the aspect of the CME  (Figure \ref{fig:Fitting_Obs}) is most easily interpreted as resulting from plasma accumulated on the surface of a horizontal torus. The clearest observations of this CME were all taken from a narrow range of helio-longitudes situated close to the Sun-Earth line in the ecliptic plane. A determination of the longitude of propagation is therefore limited in accuracy. We evaluate the impact of this uncertainty on the analysis presented here by deriving CME kinematic properties based on different assumed longitudes of propagation.\\

\begin{figure}[ht]
\centering
\includegraphics[scale=0.3]{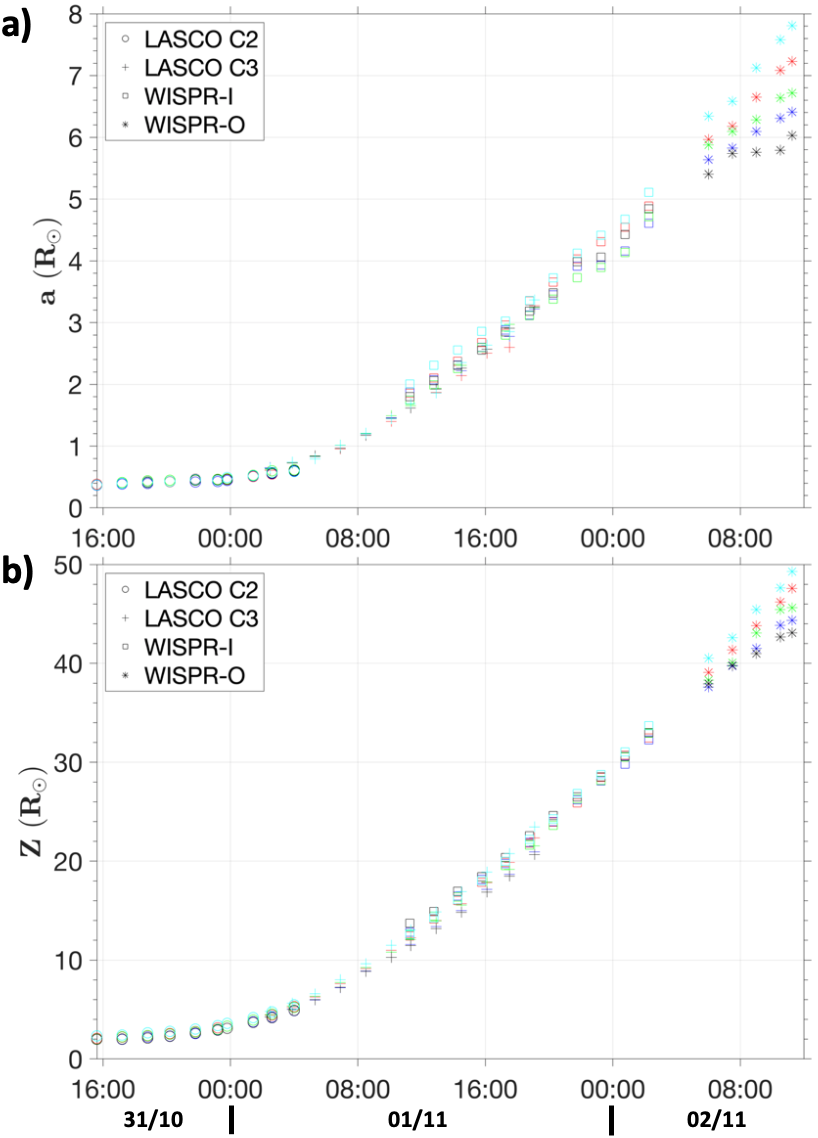}
\caption{Measurements of (a) minor radius at the apex and (b) FR height at apex from LASCO C2/C3 and WISPR I/O WL images. Different FR orientations have been assumed, and each color denotes a specific heliographic longitude: 85$^\circ$ (black), 95$^\circ$ (blue), 105$^\circ$ (green), 115$^\circ$ (red), 125$^\circ$ (cyan). \label{fig:Manual_fitting_aZ} }
\end{figure}

We fit the outline of the FR model to the bright outer boundary of the CME, indicated by the arrows in Figure \ref{fig:WL_Obs}, and do not consider the small cavity located well inside the FR \citep{Hess2019}. Figure~\ref{fig:Fitting_Obs} compares coronagraphic observations with the 3D reconstruction for a longitude of propagation of $\sim115^\circ$ in Heliographic Earth EQuatorial (HEEQ) coordinates. From such fittings, we can derive time profiles for the FR height $Z(t)$ and minor radius $a(t)$. Figure~\ref{fig:Manual_fitting_aZ} presents CME kinematics derived from reconstructions based on different longitudes of propagation. For all cases, the FR keeps the same horizontal orientation with a small tilt of $\sim$4$^\circ$ with respect to the solar equatorial plane. During the propagation, all the fits suggest that the FR is progressively deflected southward, with a latitude decreasing from $\sim$+3 to $\sim$-3 $^\circ$.\\

All reconstructions shown in Figure~\ref{fig:Manual_fitting_aZ} confirm the long eruption process discussed in \cite{Hess2019}. The derived kinematic variations are very similar inside the FOV of LASCO C2/C3, but differ at large elongation angles when the CME reaches the FOV of WISPR-O. At these large distances, the fitting becomes challenging. This is likely because \textit{PSP}'s unusual vantage point, which is situated at a smaller heliocentric distance than the CME and WISPR-O, allows it to image a CME situated further out in the heliosphere. The uncertainty in the FR position seen in Figure~\ref{fig:Manual_fitting_aZ} at these high elongation angles results from a difficult interpretation of WISPR-O images. A true multipoint observation of this CME FR would have been very helpful to reduce these uncertainties. However, an FR propagating close to the plane of the sky at a longitude of 115$^\circ$ seems to give consistent results across observing instruments at times when their FOVs overlap.\\

In the next section (\ref{sec:forces}), we describe an improved Eruptive Flux Rope (EFR) model that provides a new 3D representation of the FR magnetic field. This new model, called 3D-EFR, is developed to compute the forces acting on precisely the FR structure defined in section \ref{sec:WLRecCMEs}. We exploit this model in section \ref{sec:results_FRmodel} to study the important forces acting on the dynamic evolution of the CME imaged by \textit{SOHO} and \textit{PSP}.\\

\section{Modeling of the forces acting on the Flux Rope}
\label{sec:forces}
A 3D FR in the form of a bent cylinder, or torus, experiences toroidal forces of magnetic and plasma origins directed radially inward and outward from the center of the torus \citep{Shafranov1958,Shafranov1966}. A significant force called the 'hoop force' is induced by the poloidal magnetic field, which is stronger on the inner side than the outer side of the torus. This creates a net force that pushes the plasma torus outward and forces an expansion of the torus away from its center, shown as a black asterisk in Figure \ref{fig:shape}. The hoop force can be computed for a known poloidal field distribution via the equations of magnetostatics. All derivations start with the force balance between the Lorentz and pressure gradient force, and share the common assumption that the minor radius ($a$) of the toroid is much smaller than its major radius ($R$). For a circular current loop, such as assumed in this paper, an analytical formulation of the hoop force can be obtained by considering the self-inductance of the loop. This circumvents the logarithmic divergence encountered when integrating the radial component of the Lorentz force associated with the poloidal field. 
The resulting expression for the hoop force acting on a circular toroid with azimuthal symmetry (constant cross section) and $a/R<<1$ is:

\begin{equation}
F_H\propto I_t^2 \Bigg[ln\Bigg(\frac{8R}{a}\Bigg)-1+\frac{\xi_i}{2}\Bigg]
\label{eq:hoop}
\end{equation}
where $I_t$ is the toroidal current and $\xi_i$ is the internal self-inductance computed from the distribution of the poloidal magnetic field inside the FR. \\

The ``torus instability'' or ``lateral kink instability'', induced in part by the hoop force, can be contained in Tokamaks by imposing an additional vertical magnetic field \citep{Shafranov1966}. The latter is analogous to the ``confinement'' or ``strapping'' field in solar physics, and can correspond to solar magnetic loops overlying the FR and anchored at both ends in the dense photosphere (line-tying condition). Although the formation and eruption of magnetic FRs are time-dependent phenomena, the equations of magnetostatics have been employed in the literature to express the force balance of magnetic FRs immersed in a magnetized corona. In this approach, any imbalanced force induces an expansion or contraction of the major and minor radii of the torus. It is a powerful way to quantify the relative role of different forces on the eruption of a system. \\


A 3D FR will also experience a force induced by the toroidal component of the magnetic field. This force (the so-called ``1/R force'', $F_T$) results from the relative magnetic pressure induced by the toroidal magnetic field on the inner and outer parts of the FR \citep{Freidberg2008ppfe}. In addition, the internal plasma pressure acting on the inherently asymmetric inner and outer surface areas of the toroid exerts an additional net plasma force (the so-called ``tire-tube force'', $F_P$). 

The combined effect of the abovementioned toroidal forces ($F_H$, $F_P$ and $F_T$), the tension force of the confinement field ($F_S$), the gravitational ($F_G$) and drag ($F_D$) forces on the displacement of magnetic FRs was first solved by \cite{Chen1989ApJ,Chen1996JGR} for idealized geometry. \cite{Chen1989ApJ} assumed a modified slender toroidal structure with a varying cross section (minor radius) between the footpoints and the apex of the CME. He solved for the following equation of motion of the apex of mass $M$ at a heliocentric distance $Z$:

\begin{equation}
\label{eq:Newton}
M \frac{d^2Z}{dt^2}=F_{L}+F_P+F_G+F_D
\end{equation}

where the Lorentz forces, $F_{L}$, were decomposed into the standard three forces ($F_{H}, F_S, F_T$). These are the hoop force ($F_{H}$, see equation \ref{eq:hoop}) driven by the asymmetric distribution of the poloidal magnetic field between the inner and outer edge of the toroid, a sunward force ($F_S$) exerted by the confining coronal field, and the ``1/R force'' ($F_T$). \\

The FR aspect of (at least a subset) of CMEs inferred from \textit{SOHO} and \textit{STEREO} imaging implies that radial forces, such as the hoop force, must contribute to the strong acceleration undergone by CMEs near the Sun. This does not preclude the contribution of other effects during the formation and emergence process of the FR. Away from the Sun, the interaction of the CME with the ambient solar wind controls the kinematic properties of the CME. However, the success of the force-free field reconstructions of numerous magnetic clouds measured near 1 au suggests that Lorentz forces remain sufficiently strong to maintain the cohesion of the FR between the Sun and 1 au.
\\

The variation of the minor radius assumed in this paper (equation \ref{eq:aform}) is slightly stronger than those assumed in 3D reconstruction models of CMEs that quite successfully fit coronagraphic observations, such as the gradual cylindrical shell model of \cite{Thernisien2009}. A derivation of the hoop force (equation \ref{eq:hoop}) that includes this new variation of the minor radius can be obtained by recomputing the resulting poloidal magnetic energy in terms of revised total self-inductance ($L$):

\begin{equation}
L=4~\varphi_f \frac{R}{c^2}\Bigg[ ln\Bigg(\frac{8 R}{a}\Bigg)+\frac{1}{3}ln\Bigg(\frac{a} {a_f}\Bigg)-2+\frac{\xi_i}{2}\Bigg]
\label{eq:inductance}
\end{equation}

In this section, $a$ denotes the FR minor radius at apex $a_a=a(\varphi=0)$. For the initial condition $a=a_f$, the term $\frac{1}{3}ln(a/a_f)$ is zero and we retrieve the total self-inductance of a toroid with constant minor radius. In contrast to previous expressions for $a$, the new FR geometry can be used to define a fully 3D magnetic field inside the FR.\\

The kinematic model of the FR used in this study follows the calculation of \cite{Chen1989ApJ,Chen1996JGR} by integrating the force balance equation (equation \ref{eq:Newton}) along the apex of the CME. Projecting magnetic and plasma forces along the radial direction, the equation of motion takes the following well-known form:

        \small
		\begin{equation}
		\label{eq:forcebalance}
		\begin{aligned}
			F_R(l) &= \underbrace{\frac{I^2_t}{c^2R}}_{K}\left[\underbrace{ln\left(\frac{8R}{a}\right)-1+\frac{l_i}{2}}_{\equiv F_H} \underbrace{+\frac{1}{2}\beta_p}_{\equiv F_P} \underbrace{-\frac{1}{2}\frac{\bar{B_{t0}}^2}{B^2_{pa}}}_{\equiv F_T} \underbrace{+2\left(\frac{R}{a} \right)\frac{B_s}{B_{pa}}}_{\equiv F_S} \right] \\
			&+ F_G + F_D \\
	    \end{aligned}
		\end{equation}
        \normalsize

where $\beta_p=8\pi(\bar{P}-P_a)/B^2_{pa}$ is the plasma beta parameter, $\bar{P}$ is the average pressure inside the FR, $P_a$ is the ambient coronal pressure, $\bar{B_{t0}}$ is the average zeroth-order toroidal magnetic field inside the FR, $B_{pa}=Bp(r=a)=2I_t/(ca)$, $l_i = \frac{2}{a^2 B^2_{pa}} \int_0^a B^2_{p}(r) r dr$ is the internal inductance, and $I_t=2\pi\int J_t(r)rdr=\Phi_p/(cL)$ is the total toroidal current ($L$ is the effective loop inductance). Here, $\Phi_P$ is the total poloidal flux computed at the apex of the FR.\newline 

In addition to changing the effective inductance to accommodate the assumed new geometry of the FR (equation \ref{eq:aform}) in the computation of the Lorentz force, we also change the form of the background magnetic field ($B_s$). This confining field was calculated in previous studies by assuming that it is always parallel to the poloidal component of the FR \citep{Chen1996JGR}. This field induces a sunward-directed force that can counteract the effect of the radial forces (i.e. via $F_S$), including the hoop force. In 3D-EFR, the confinement magnetic field is obtained directly from a Potential Field Source Surface (PFSS) model \citep{Wang1992}. This further limits the number of free parameters of the model and provides a more realistic description of the background corona than assumed in previous applications of the EFR model. The component of the background coronal field parallel to the FR poloidal field is computed dynamically from the PFSS model as the FR rises in the atmosphere. For this study, we based the PFSS extrapolation on magnetograms provided by the Wilcox Solar Observatory (WSO).\\

Momentum coupling of the FR with the ambient solar wind can either slow down a fast FR propagating in slower wind or accelerate a slow FR pushed by faster wind. The drag force ($F_D$) in 3D-EFR is expressed as:
\begin{equation}
\label{eq:Drag}
F_d=c_dn_am_ia (V_{SW}-V)\left|V_{SW}-V\right|
\end{equation}
where $n_a$ is the ambient density, $m_i$ the ion mass, $V=dZ/dt$ is the speed of the FR apex, $V_{SW}(Z)$ is the speed of the ambient solar wind at the leading edge of the CME, and $c_d=1$ is the dimensionless drag coefficient. This expression of $F_d$ assumes that the Reynolds number is high and that turbulent flows develop around the FR. The drag force develops when the FR exits the loops of the helmet streamers and enters a region dominated by the outflowing solar wind. The force increases with the difference in speed between the FR and the background solar wind $V_{SW}$. The FR studied here propagates in the slow solar wind above helmet streamers. Therefore, in this study, we use a background solar wind profile derived from measurements of densities fluctuations along streamers stalks \citep{Sheeley1997ApJ,Sanchez2017ApJ}:

\begin{equation}
\label{eq:solarwind}
V_{SW} = 190 [\tanh{[3\times10^{-7}(Z-4\times10^6) ]} + 1] -75
\end{equation}
with $V_{SW}$ in km/s and $Z$ in km. We will show that the drag force can play an important role in the acceleration of the CME studied in this paper.\\

The minor radius $a(t)$ of the FR is changed in time according to the following differential equation also taken from \cite{Chen1996JGR}:

		\begin{equation}
		\label{eq:Minor_direction_equation}
			M\frac{dw}{dt}=\frac{I_t^2}{c^2 a}\left[\frac{\bar{B_{t0}}^2}{B^2_{pa}} -1 + \beta_p\right]
		\end{equation}

where $w=da/dt$ is the minor radial growth speed, and $\beta_p=8\pi(\bar{P}-P_a)/B^2_{pa}$ is again the plasma beta parameter. The size of the cross section is therefore controlled by the contracting effects of the poloidal field, the expanding effects of the axial field, and the pressure gradient between the inside and outside of the FR. 


\cite{Chen1989ApJ} solved for the dynamic coupling between the force-balance equation of the FR motion (equation \ref{eq:Newton}) and the expansion of the minor radius (equation \ref{eq:Minor_direction_equation}). These coupled equations constitute a complete semi-analytical treatment of the apex of the FR. \\

In Appendix A, we derive from the Grad-Shafranov equation (see \cite{Shafranov1966} and in \cite{Priest2014masu}) analytical  expressions for the 3D internal magnetic field structure of the FR. These calculations assume axi-symmetric magnetic fields (i.e. without dependence in the azimuthal $\varphi$ angle), such that the FRs has both uniform major ($R$) and minor radii $a$. The functional form assumed for $a(\varphi)$ (equation \ref{eq:aform}) is such that $a(\varphi)$ does not vary significantly over an angular extent of $45^\circ$ on either side of the apex (i.e. in the range of $\varphi=-45^\circ$ to $45^\circ$). Toroidal symmetry is therefore roughly fulfilled for a broad region near the apex of the FR, but is not down the legs of the FR. To derive a magnetic field distribution, we use solutions of the 2D Grad-Shafranov equation (derived in Appendix A) for 100 cross sections (or equivalently, 100 $\varphi$ angles) of the FR all along the toroidal axis. We then consider all solutions along the toroidal axis and reconstruct the global 3D magnetic field lines.
\\

In summary, the model is run as follows. We assume that an initial FR already exists prior to the eruption. We define an equilibrium condition that depends on a specified initial height $Z_0$, footpoint separation ($S_f$), and aspect ratio ($a/R$), as well as densities and temperatures inside and outside the FR. These input parameters are the same as in \cite{Chen1996JGR}. Setting equations \ref{eq:forcebalance} and \ref{eq:Minor_direction_equation} to zero provides the initial value of the poloidal field for an initial confinement field given by the PFSS coronal model. As in \cite{Chen1996JGR}, the structure is destabilized by increasing the amount of poloidal magnetic flux ($\Phi_P$) of the FR, which produces a stronger set of forces forcing the CME to erupt. Numerical integration of equations \ref{eq:forcebalance} and \ref{eq:Minor_direction_equation} provides the evolution of the minor ($a$) and major ($R$) radii of the FR, as well as the evolution of the maximum toroidal and poloidal magnetic fields. From these values, we then assume that the zeroth-order toroidal magnetic field follows equation \ref{eq:BT0_profile}, and we solve the Grad-Shafranov equation to obtain a full description of the internal magnetic field and its first-order asymmetries given by equations \ref{Inside_field_equation}.\\
We now exploit 3D-EFR to interpret the eruption, the propagation, and the morphology of the CME imaged by \textit{SOHO} and \textit{PSP}.
\vspace{1cm}

\section{Modeling the CME imaged by WISPR} 
\label{sec:results_FRmodel}

\citet{Hess2019} shows that a coherent structure has already formed in the low corona. Unfortunately, we do not have spectropolarimetric observations of that cavity that could have provided additional clues on the 3D topology of the magnetic field by using data-optimized FR modeling techniques \citep{Dalmasse2019ApJ}. We assume that an FR already exist inside this cavity. The 3D reconstruction carried out in section \ref{sec:Manual_tracking} provides the direction of propagation, namely a HEEQ longitude of 115$^\circ$, a heliographic latitude that changes progressively from $\sim 3^\circ$ at onset to -3$^\circ$ in WISPR-O images, and a tilt angle of the modeled FR of 4$^\circ$. The initial height of the FR is set at 0.5 R$_\odot$, just above the outer edge of the AIA field of view at 0.43 R$_\odot$, at a height where the cavity becomes less deflected in latitude by the ambient coronal magnetic field \citep{Hess2019}. Past studies of EUV cavities reveal that their densities are typically 70-80\% that of the surrounding streamer material \citep{Schmit2011ApJ}. While some cavities tend to have temperatures similar to those of their surrounding media, others appear hotter \citep{Gibson2018}. We begin by using the properties of the cavity inferred by \citet{Hess2019} with a temperature of about 1 MK and a density equal to 70\% of the ambient streamer material. We also briefly discuss the results of running 3D-EFR for a hotter cavity with a stronger density depletion. \\

We drive the cavity eruption from the inner corona by an enhancement of the poloidal flux that induces a weak hoop force pushing the structure very gradually out of the corona. We will later consider the possible effect of buoyancy acting on this initial eruption. The second, more pronounced acceleration of the FR, at a heliocentric radial distance of 2-6 R$_\circ$, occurs where the CME exits the helmet streamer and enters the open magnetic field of the solar wind. Past surveys of CMEs that accelerated strongly near 2-6 R$_\circ$ in the LASCO coronagraphs have shown that their releases are frequently associated with material also moving sunward \citep{Wang2006ApJ}. In these events, the outward component is shaped like a large arch with both ends attached to the Sun, and the inward component ('inflows') consists of collapsing loop-like structures \citep{Wang2006ApJ}. These observations are interpreted as the effect of magnetic reconnection adding helical magnetic fields to the CME, and a byproduct of this is a system of arcades collapsing sunward \citep{Sheeley2007ApJ}. LASCO C2 did not detect inflows for the event analyzed in the present study. This could point to a rather weak reconfiguration of the CME topology, or it may indicate that inflows caused brightness variations that were below the detection levels of the LASCO instrument. One possibility is therefore that a weak reconfiguration of the magnetic field occurs above the tip of the streamers that increases the poloidal magnetic field pushing the FR outward. Full 3D MHD simulations of weak and slow CMEs suggest that the momentum coupling of the background slow solar wind could also contribute greatly to the acceleration of the CMEs \citep{Lynch2016}. The drag between the FR and the slow wind could explain the fact that slow CMEs move at the speed of the ambient slow wind \citep{Lynch2016}.\\

\begin{figure}[h!]
\centering
\includegraphics[scale=0.28]{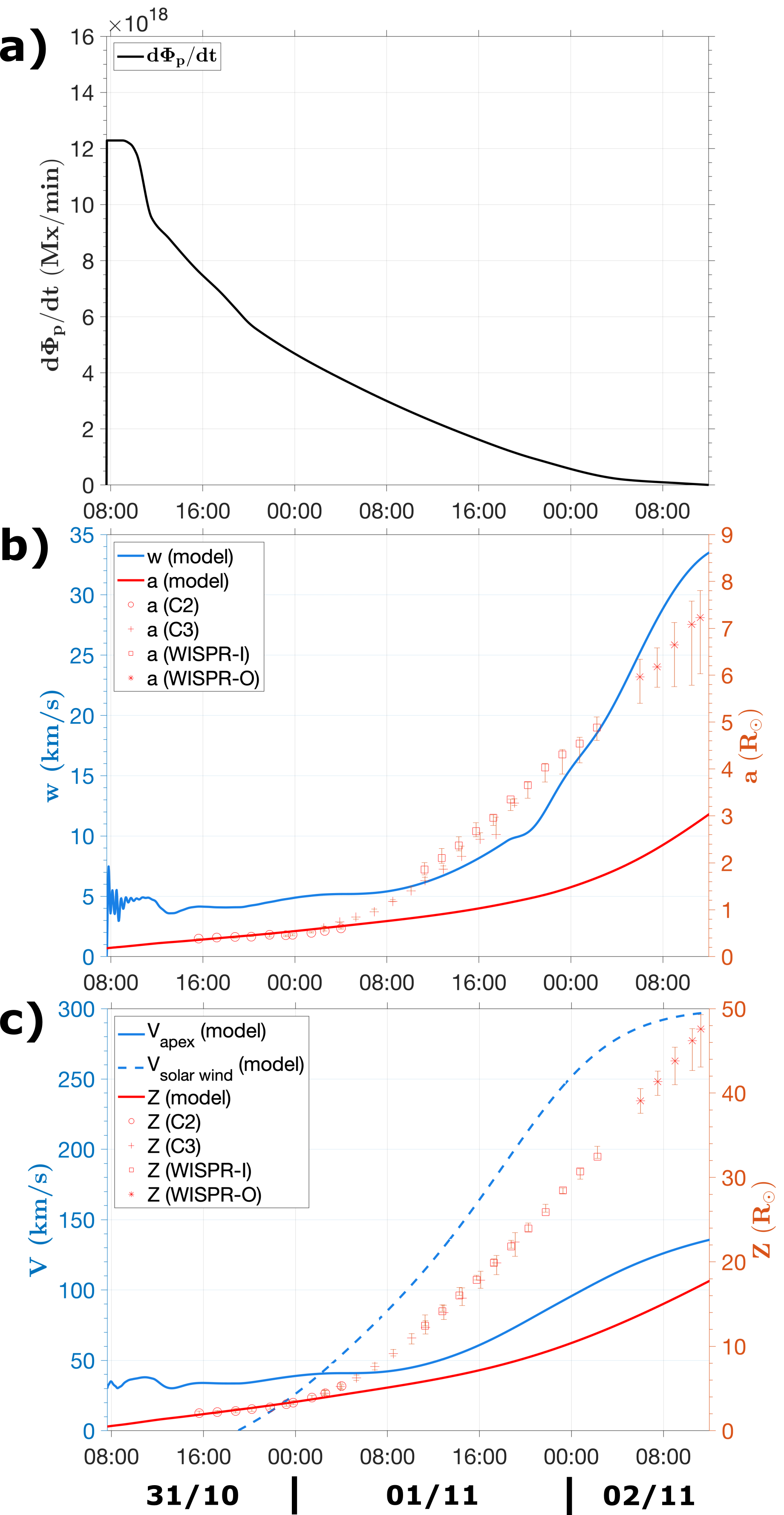}
\caption{Panel (a) shows the poloidal flux injection rate ($d\phi_p/dt$). Panel (b) displays the minor radius ($a$) of the FR at apex (in red) and expansion rate ($w=da/dt$) (in blue). Panel (c) illustrates the time evolution of the apex height ($Z$) (in red) and apex speed ($V=dZ/dt$) (in blue) of the FR. The ambient solar wind speed (from equation \ref{eq:solarwind}) assumed in the model is plotted as a dashed blue line. Panels (b) and (c) show the results of the 3D reconstruction (Figure~\ref{fig:Manual_fitting_aZ}), plotted with markers and error bars. The error bars correspond to the results dispersion induced by the different assumed orientations (see Section~\ref{sec:Manual_tracking}). \label{fig:Fitted_Kinematics} }
\end{figure}

Therefore, we study the relative contribution of both the hoop force and the drag force on the second acceleration. We first test whether the drag force can drive the second acceleration of the FR in the region where the slow wind accelerates. For that, we prescribe a  poloidal flux injection that peaks in the low corona to drive the cavity motion toward the outer corona and decrease the injection rate gradually as the CME passes 3-5 R$_\odot$. The results are shown in Figure~\ref{fig:Fitted_Kinematics} as a function of time. The poloidal flux injection rate ($d\Phi_P /dt$, panel a) is compared with the minor radius expansion rate ($w(t)=da/dt$, panel b) and apex speed ($V(t)=dZ/dt$, panel c). The flux injection rate (panel a) peaks between 07:40 and 10:00 and 23:00UT on 2018 October 31, and then decreases gradually to zero. Data points from the 3D reconstruction shown in Figure~\ref{fig:Manual_fitting_aZ} are also plotted as red circles and stars in Figure \ref{fig:Fitted_Kinematics}a and \ref{fig:Fitted_Kinematics}b. \\

The oscillations visible for $w$ (panel b) and slightly for $V$ (panel c) are induced by the sudden variations enforced on $d\Phi_P /dt$ in order to initiate the FR propagation. The speed of the FR (blue line, panel c) increases in response to the increasing solar wind speed (dashed blue line, panel c) and the associated effect of the drag force. However, we find that the drag, while a significant contributor to the acceleration of the FR, appears insufficient to reproduce the terminal speed of the CME (300 km/s). In this run, the terminal speed reaches about 130 km/s, which is less than half of the CME terminal speed of ~300 km/s derived from observations (Figure \ref{fig:Manual_fitting_aZ}). In addition, without a second injection of poloidal flux, the internal magnetic field remains weak and the minor radius $a$ too small. Propagating the CME all the way to 1 au, we find that the magnetic field signature is not representative of the field strength that we typically measure in slow CMEs at 1AU. This point is addressed further in the discussion section. \\

In order to reproduce the CME dynamics inferred in Figure \ref{fig:Manual_fitting_aZ}, a second injection of poloidal flux seems necessary, in order to significantly boost the hoop force in that region. As already discussed, magnetic reconnection is one possible mechanism that would force ambient coronal loops to merge and produce a second enhancement of the internal poloidal flux, but other mechanisms are also addressed in the discussion section.  \\


\begin{figure}[h!]
\centering
\includegraphics[scale=0.28]{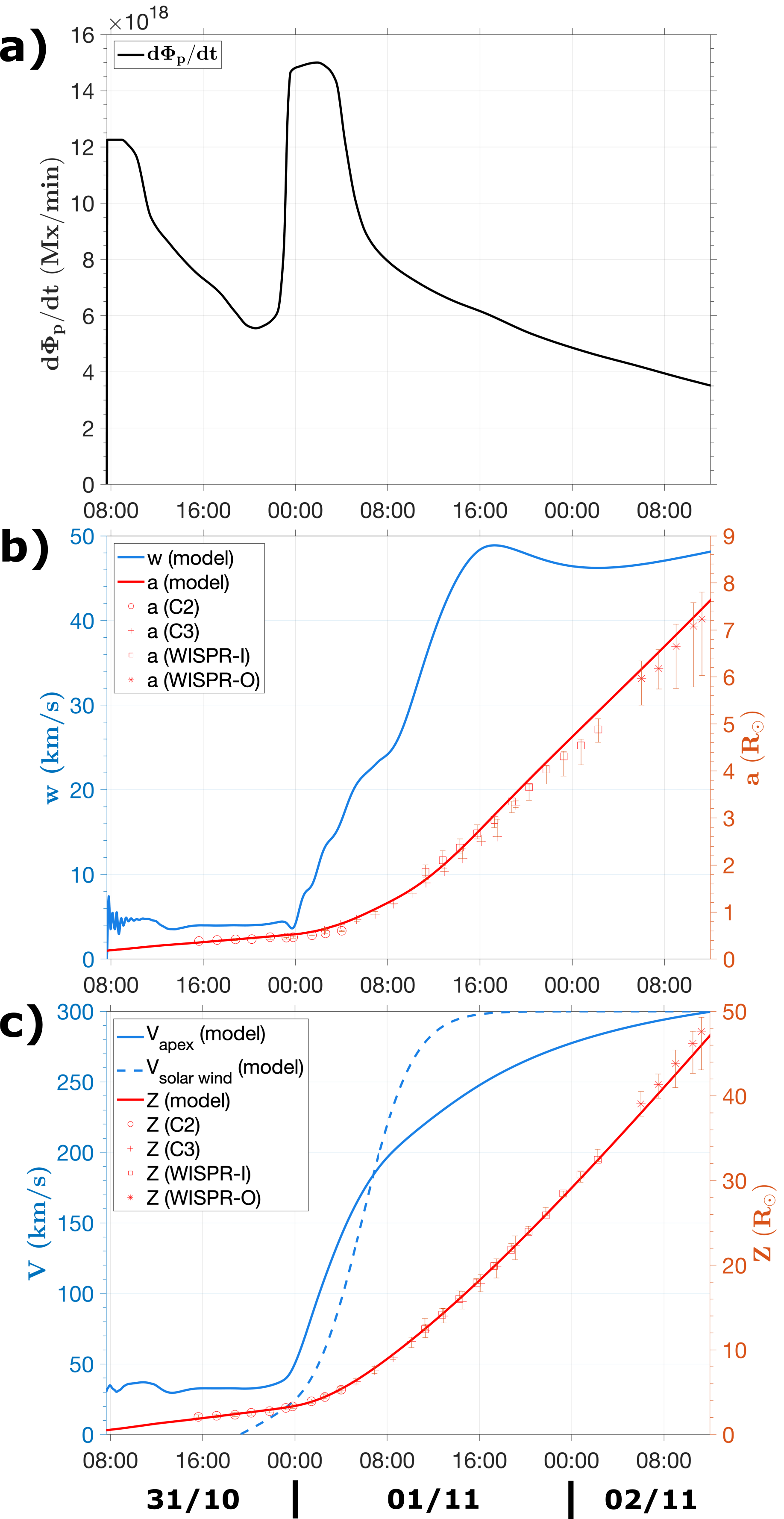}
\caption{Same format as Figure \ref{fig:Fitted_Kinematics}, but for a simulation result that implements two phases of poloidal flux injection in order to reproduce the observed kinematics.} \label{fig:Fitted_Kinematics_twoinj} 
\end{figure}

We therefore compute the poloidal magnetic flux and the associated hoop force necessary to reproduce the observed kinematic properties. The results are shown in Figure \ref{fig:Fitted_Kinematics_twoinj}. The two flux injections are seen in panel (a) between 07:40 and 23:00UT on 2018 October 31, and the larger peak between 00:00 and 06:00UT on 2018 November 1. The first injection leads again to a gradual motion of the FR from the low to the high corona, as well as a weak expansion rate of the minor radius. The second injection induces a strong acceleration of the FR, with speeds increasing from less than 50 $km/s$ to greater than 250 $km/s$. During that latter phase the minor radius increases suddenly. The figure shows that this run of 3D-EFR reproduces very well the evolution of the FR apex height and the minor radius in all fields of views. \\ 

\begin{figure*}[ht]
\centering
\includegraphics[scale=0.3]{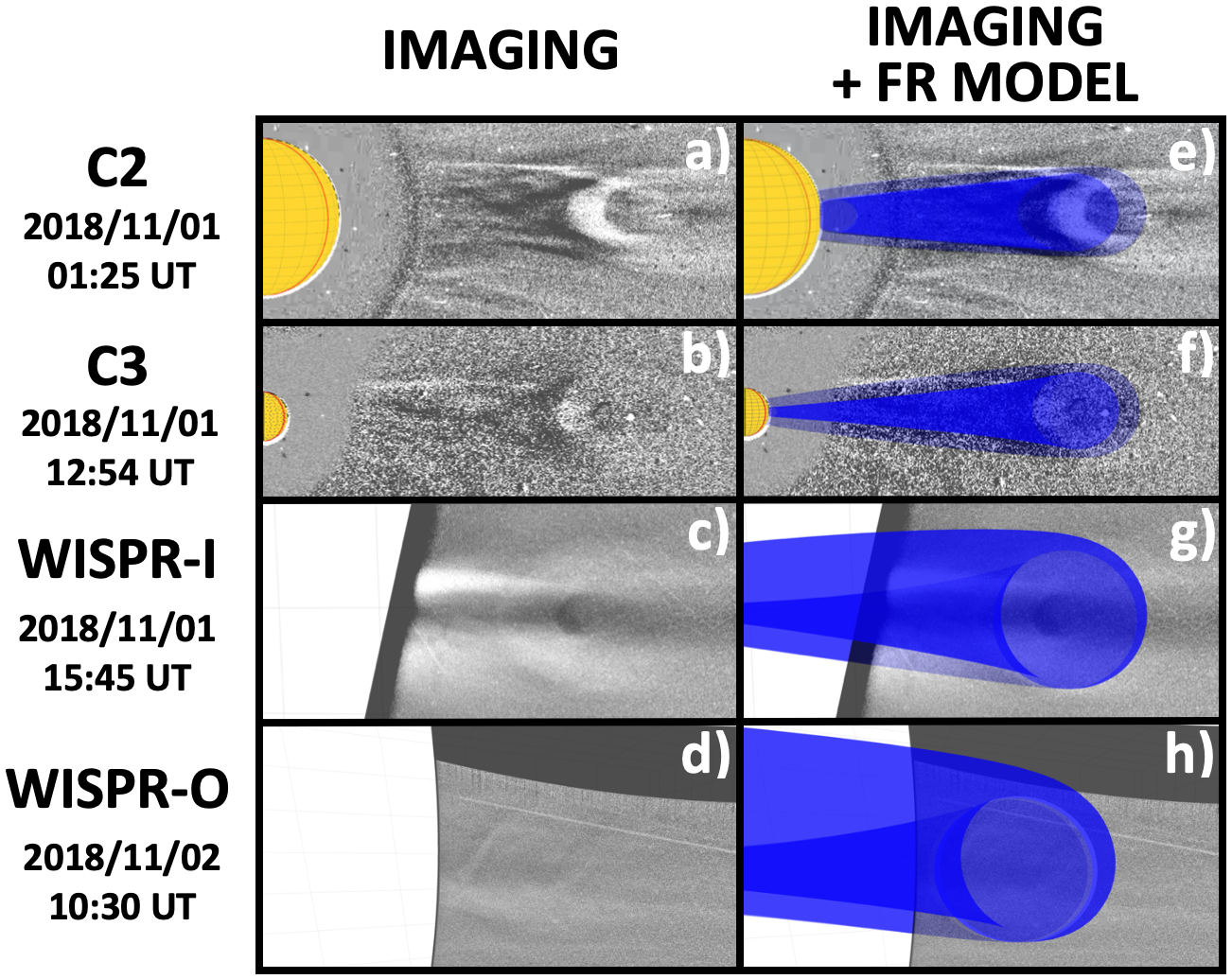}
\caption{The same as in Figure~\ref{fig:Fitting_Obs} but with the kinematics given by the eruptive FR model shown in Figure~\ref{fig:Fitted_Kinematics_twoinj}. \label{fig:FRmodel_Fitting_Obs} }
\end{figure*}

Figure \ref{fig:FRmodel_Fitting_Obs} presents a comparison of the modeled FR with two flux injections (Figure \ref{fig:Fitted_Kinematics_twoinj}) with the white-light observations. We use the same representation as the one comparing observations with the 3D reconstruction technique presented in Section~\ref{sec:Manual_tracking}. The aspect of the modeled FR surface is compared with the running-difference images of C2 (panels a and e), C3 (panels b and f), and Level-3 images of WISPR-I (panels c and g) and WISPR-O (panels d and h) already shown in Figure \ref{fig:Fitting_Obs}. Overall, there is a good agreement between the modeled and observed CME, except in WISPR-O, where multiple fronts are observed that are not explained by the model. Moreover, the modeled FR appears larger in WISPR-O. This is an issue with the 3D interpretation that we already discussed in Figure~\ref{fig:Fitting_Obs}.\\

The part of the CME that is most clearly imaged in LASCO is its back end, where an outward-moving concave structure develops into the brightest feature imaged during this event. This concave structure is very common in slow CME events, and can become the dominant feature observed in white light \citep{Sheeley2007ApJ}. The concave shape has been associated with the sunward surface of magnetic flux ropes \citep{Thernisien2009}, but we discuss alternative interpretations in the discussion section. Tracking of these concave structures to spacecraft making in-situ measurements shows a clear association between their passage and the time when the spacecraft exits the poloidal magnetic field situated on the sunward edge (back end) of the magnetic flux rope \citep{Moestl2009,Rouillard2009a}. The dark circular feature observed in the WISPR-I image (panel c) is situated well inside the surface of the FR, closer to its current channel.    \\


\begin{figure}[ht!]
\centering
\includegraphics[scale=0.28]{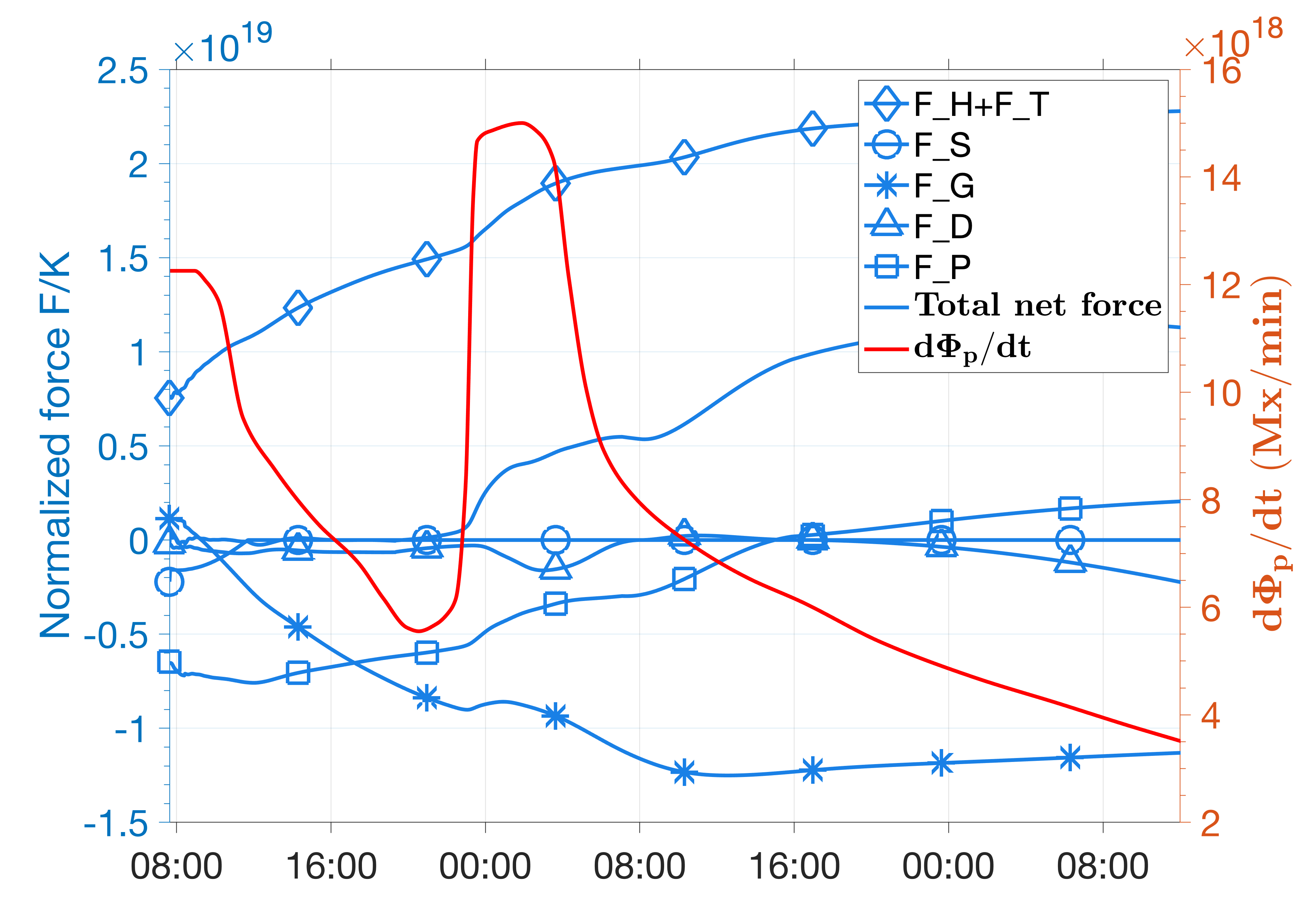}
\caption{Time evolution of the radial components of forces acting on the apex of the magnetic FR as a function of time. The forces shown are the toroidal forces ($F_H$, $F_P$, and $F_T$), the tension force of the confinement field ($F_S$), and the gravitational ($F_G$) and drag ($F_D$) forces. Negative values represent sunward-pointing values, while positive values are antisunward-pointing. Forces are normalized by $K=I_t^2/(c^2 R)$. The poloidal flux injection rate is shown as the red curve, and its values are given along the right-hand ordinate.\label{fig:CHEN_forces} }
\end{figure}

Figure \ref{fig:CHEN_forces} presents the evolution of the magnitude of the forces acting on the modeled FR when we implement two poloidal flux injections (Figure \ref{fig:Fitted_Kinematics_twoinj}). Overall, the dominant outward-pointing forces are the combined hoop ($F_H$) and $1/R$ ($F_T$) forces (diamond markers) in the case of the two injections. These two forces act to accelerate the structure out of the corona. The first episode of flux injection that lasts until 23:00 UT on 2018 October 31 is associated with a gradual increase of these combined forces. The confinement force ($F_S$, circle markers) is computed from the value and orientation of the background magnetic field given by PFSS at each location of the FR. This force initially acts to limit the acceleration in the very low corona, but becomes negligible by 13:00 UT. The second episode of poloidal flux injection boosts the hoop force and the FR speed, which limits greatly the role of the drag force ($F_D$), because the difference in speed between the FR and the wind becomes much smaller.  \\

Instead of the cavity properties inferred from the analysis of \citet{Hess2019}, i.e. a cold cavity of $\lesssim$1 MK with 20\% density depletion relative to the ambient medium, we run 3D-EFR for a cavity temperature to 3 MK with a stronger 80\% density depletion. We noticed some interesting differences. Low in the corona, the greater density depletion leads to a stronger buoyancy force ($F_G$) comparable in magnitude to the rather weak hoop and $1/R$ forces. Higher up, in the WISPR-I field of view, the minor radius and kinematics of the FR could not be fitted as well to the observations.  \\

\begin{figure}[ht!]
\centering
\includegraphics[scale=0.24]{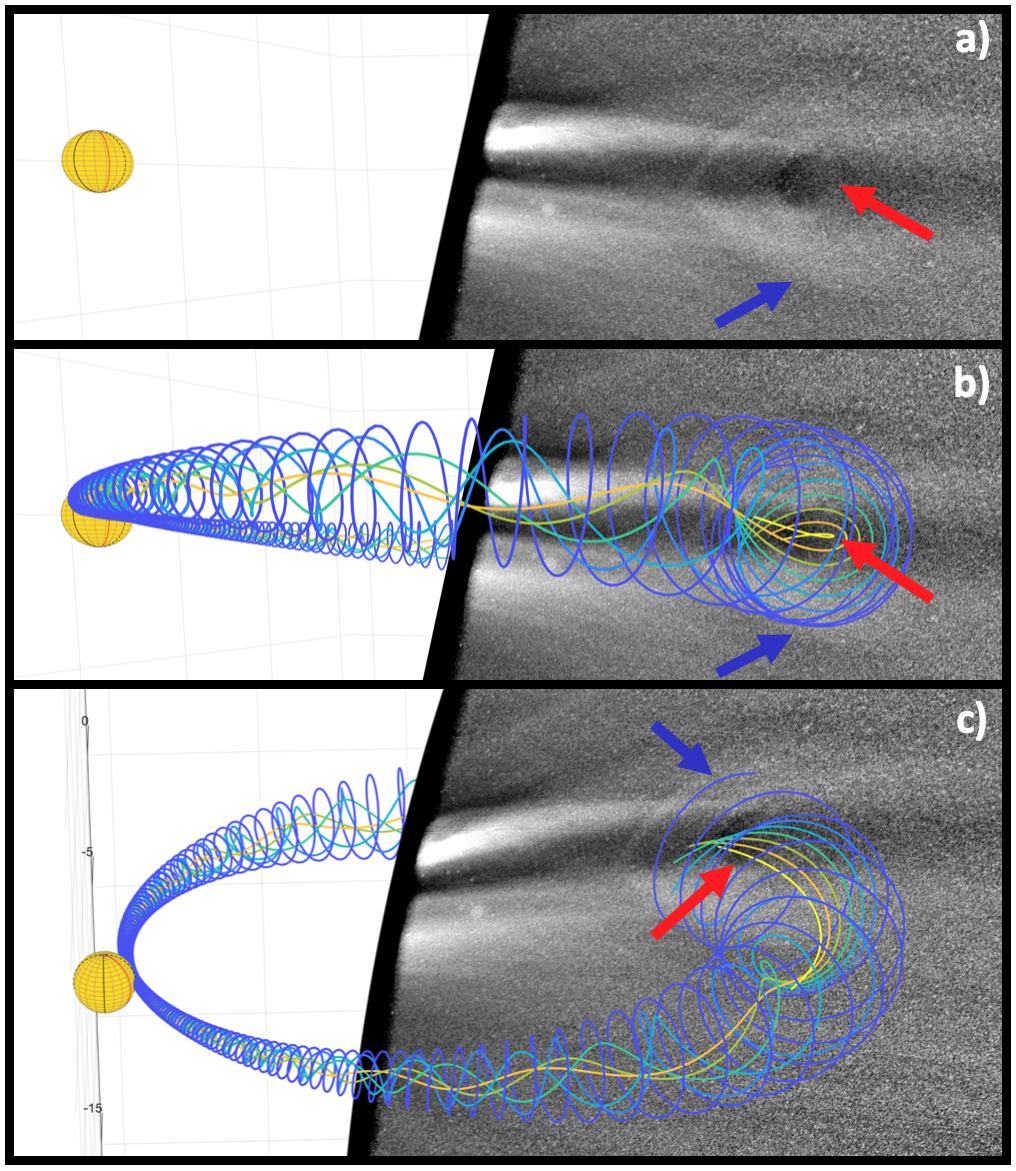}
\caption{Panel (a): an image from WISPR-I taken on 2018 November 1 at 19:30:50 UT. Panel (b): the same image as in panel (a) but with the results of the 3D flux rope fit superimposed. Panel (c): the same as in panel (b) but from another viewpoint than \textit{PSP}. The magnetic field lines computed by the model presented in this paper are traced inside the FR. The bright ring (blue arrow) corresponds to plasma located at the boundary of the FR where the poloidal magnetic field dominates. The dark core (red arrow) marks the location where strong axial magnetic fields (yellow lines) dominate the plasma locally. \label{fig:FR_FLs} }
\end{figure}

The kinematics and size of the CME being represented relatively well by the 3D-EFR run with two flux injections, we next consider the relation between the images and the internal 3D magnetic field of the FR. Figure \ref{fig:FR_FLs} provides a comparison between the magnetic field structure of the modeled FR and the white-light images. The dark blue and yellow lines depict magnetic field lines that have strong poloidal and toroidal components, respectively. As discussed before, the outer extent of the CME observed in white-light images is associated with the regions of the flux rope where the poloidal magnetic field dominates. In contrast, the circular dark region visible in the images corresponds fairly well to the region where there is mostly axial (i.e. toroidal) magnetic field.   \\

\section{Discussion} 

The 3D-EFR model presented is a modification of the \citet{Chen1996JGR} model that computes, from the magnetostatic equations, the toroidal forces acting on a slender flux rope. The main goals of this significant upgrade were to model FR structures more realistically and to decrease the number of free parameters. The 3D-EFR model implements different modules to compare the 3D geometry of the FR with observations (Figure \ref{fig:FRmodel_Fitting_Obs}). The basic modifications made to the model are summarized below. \\

\begin{itemize}
    \item The model assumes a new variation of the minor radius of the FR, from its footpoint to its apex (equation \ref{eq:aform}), that removes discontinuities in flux surfaces that previously prevented a truly 3D description of the magnetic field. The new inductance derived from the form of $a$ is given in equation \ref{eq:inductance}. 
    
    \item The model exploits a more realistic description of the coronal magnetic field based on PFSS to calculate the confinement force as the FR progresses from the low to the upper corona. In future developments, we will also exploit nonlinear force-free field extrapolations as well as 3D MHD models that account for the presence of currents near the source regions of more energetic CMEs.
    
    \item The model incorporates a description of the FR magnetic field in 3D by solving the Grad-Shafranov equation along 2D cross sections of the FR (equation \ref{eq:B_components_equation}). In future studies, we will exploit this description to investigate how magnetic flux surfaces are shifted by the effect of the Lorentz force in more powerful CMEs. 
\end{itemize}

A first application of the model was to investigate the physical mechanisms that could lead to the CME eruption in two phases. \\

We show that the eruption of the cavity from the low corona and its motion to the upper corona (3-4 R$_\odot$) could be driven by a small enhancement of the internal magnetic field. This creates a weak hoop force that drives the cavity'’s motion to the upper corona. For the cavity properties derived by \citet{Hess2019}, we find that the buoyancy force does not contribute significantly to this motion. We show that this force could be more significant for cavities with lower densities. \\

We also investigated the physical mechanisms driving the second, more pronounced acceleration of the CME near 3-5 R$_\odot$. Because the slow solar wind accelerates in this region, a natural mechanism to first investigate was momentum coupling induced by the slow wind and the FR. We found that this drag force can account for half the acceleration experienced by the FR when no additional poloidal flux is injected. Our treatment of the drag force is, however, very simple. The form of the turbulent flows that form in the wind as it deflects around the backend of the CME may affect the properties of the drag.  These effects should be investigated more thoroughly using high-resolution 3D MHD simulations.\\

A remarkable match between the modeled kinematic evolution and expansion of the FR and the observations is obtained when we include a second injection of poloidal flux when the FR reaches 3-4 R$_\odot$. This injection boosts the hoop force that accelerates the FR and regulates the size of the cross section. The increase in CME speed at this height decreases the speed difference between the FR and the ambient wind, and therefore limits the influence of the drag force.\\

\citet{Vrsnak2019} investigated three physical processes that could induce the gradual rise phase of FRs in the corona, but for the case of a much faster CME than the event considered in this paper. He investigated the effect of a twisting motion at the FR footpoint, the emergence of new magnetic flux beneath the FR, and mass leakage down the FR legs. He concluded that the enhancement of the FR electric current, the increase of the twist, and the mass loss are tightly related phenomena, expected to occur jointly during the gradual pre-eruptive phase of an eruption. The conclusions of the present study agree with the conclusion of \citet{Vrsnak2019} that increasing gradually the poloidal field of the FR can cause its slow motion to the upper corona. We have tentatively related the origin of this increase to magnetic reconnection progressively adding magnetic flux at the back of the FR \citep[as in, e.g.,][]{Aulanier2012}. \\

Unfortunately, this CME did not cross the \textit{PSP} trajectory despite passing very close to the spacecraft. Having in situ data for this event would have been extremely helpful to better constrain the model parameters. We can, however, compare the properties of the modeled FR with the magnetic fields typically measured in situ during slow Interplanetary CMEs (ICMEs). In rare cases where such slow streamer CMEs ($<$400 $km/s$) have been tracked continuously all the way to 1 au, it was found that the maximum values of the internal magnetic field are typically in the range of 10\,--\,20 nT \citep{Moestl2009, Rouillard2009a,Rouillard2009b}. We ran the modeled FR with the two poloidal flux injections all the way to 1 au, and found that the magnetic field strength inside the FR is about 12 nT at 1AU. We therefore conclude that the amount of poloidal magnetic flux injected in the FR in this study is reasonable.\\

A slow CME erupted several days after the event presented in this paper, and was measured in situ by \textit{PSP} as a magnetic cloud on 2018 November 12 by \textit{PSP} \citep{Korreck2020ApJS}. The maximum strength of the magnetic field measured in situ was $\sim$100 nT when it passed by \textit{PSP} at a heliocentric radial distance of 55 R$_\odot$. The second acceleration of that CME occurred even higher up in the corona, near 8-10 R$_\odot$, from less than 100 km/s near 18UT on November 10 to over 350 km/s when it exited the COR-2A field of view at around 6UT on November 11 near 19 R$_\odot$ \citep{Comas2019Nat,Korreck2020ApJS,Nieves2020ApJS}. For the event considered in this paper, we derived a magnetic field magnitude of about $\sim$ 250 nT, which is higher than the CME measured on 2018 November 12 by \textit{PSP}. An interpretation could reside in the evolution of the two CMEs in white-light images. The 2018 November 1 CME, analyzed here, accelerated to supersonic speeds near 3-5 R$_\odot$, in contrast to the CME that impacted \textit{PSP} on 2018 November 12, which accelerated to high speeds near 6-8 R$_\odot$. If the acceleration is induced by a reconfiguration of the coronal magnetic field, as suggested in the present paper, then we should expect the November 12 CME measured by \textit{PSP} to have formed in weaker magnetic fields than the CME considered here. This could be the reason the internal magnetic field of the 2018 November 12 CME is weaker than for the event studied here. \\

Comparison of the model with the white-light images showed that the regions of the FR where the poloidal component of the magnetic field dominates are brighter than the cavity of the CME. This could result from four possible effects. First, as already stated, the brightness of the concave structure at the back-end of the FR could result from the horizontal orientation of the FR \citep{Thernisien2009,Rouillard2009a}. Second, the poloidal field component of the FR is here interpreted as the result of magnetic reconnection between streamer loops. The high-density plasma on these loops must be transferred to the helical magnetic field lines situated on the periphery of the flux rope. This would enhance the brightness of the poloidal field on the periphery. Third, the magnetic reconnection of field lines during the pinch-off occurring at the back end of the CME during the fast eruption creates kinks in the field lines that must be attenuated by the tension force. This produces an acceleration of the field line toward the center of the FR, and likely an enhancement of plasma density in this region due to the field lines sweeping the plasma. Fourth, if the acceleration of the CME were driven by momentum coupling, then the interaction of the accelerating solar wind with the back end of the CME would also enhance density locally. These relative processes should be analyzed in a future study. \\

\section{Conclusion} 
The analysis presented here was limited to one CME imaged clearly by WISPR. The model should be applied to more cases of similarly slow CME events that have been imaged and measured in situ by \textit{SOHO}, \textit{SDO} and \textit{STEREO}. Future applications of the model will also consider faster and more impulsive CMEs that typically accelerate lower in the corona from regions with stronger magnetic fields. For these events, the present model allows us to study how the internal magnetic field structure is deformed by the Lorentz force. The model presented here runs in seconds, and therefore offers interesting space-weather capabilities. This will be investigated in future studies.\\

We note that the present model ignores the deformation of the FR due to its interaction with the solar wind plasma. Such interactions can result from the compression of the slow solar wind by fast CMEs \citep[e.g.][]{Temmer2011ApJ} or from the compression of slow CMEs by high-speed streams \citep[e.g.][]{Rouillard2010}. Both scenarios can cause important geomagnetic storms, depending on whether the compressed part of the flux rope contains south-pointing magnetic fields or not \citep{Fenrich1998}. A procedure to model these deformations by some form of simple parameterization could be highly beneficial to improve the space-weather capabilities of the model.\\

We have also adapted our software to include the orbit of the \textit{Solar Orbiter} and the images that will be acquired by that mission in the near future \citep[see review paper by][]{Rouillard2019}. We hope that, in future studies, we will be able to combine data from \textit{PSP} and \textit{Solar Orbiter} to track and model the evolution of CMEs from their birth near the Sun to Earth-like distances. As \textit{PSP} gets closer to the Sun, we will be able to study the internal magnetic field of the CME in regions where it still accelerates strongly. These measurements will provide new information on the relative forces acting on CMEs. The 3D-EFR model will be soon available to run via a web-based interface written in Java at \url{http://spaceweathertool.cdpp.eu/}. A publication dedicated to the presentation of this interface will be submitted in the near future.

\acknowledgments
We are grateful to Dr J. Chen for many constructive discussions on magnetic flux ropes, as well as for allowing us to compare his program of eruptive flux rope model with the one developed in the present paper. The IRAP team acknowledges support from the French space agency (Centre National des Etudes Spatiales; CNES; \url{https://cnes.fr/fr}), which funds activity in the Plasma Physics Data Center (Centre de Données de la Physique des Plasmas; CDPP; \url{http://cdpp.eu/}), and the Multi Experiment Data \& Operation Center (MEDOC; \url{https://idoc.ias.u-psud.fr/MEDOC}), as well as the space-weather team in Toulouse (Solar-Terrestrial Observations and Modelling Service; STORMS; \url{https://stormsweb.irap.omp.eu/}). This includes funding for the data mining tools AMDA (\url{http://amda.cdpp.eu/}), CLWEB (\url{clweb.cesr.fr/}), and the propagation tool (\url{http://propagationtool.cdpp.eu}). The work of A.P.R. and N.P. was funded by the ERC SLOW{\_}\,SOURCE project (SLOW{\_}\,SOURCE - DLV-819189). A.K. also acknowledges financial support from the COROSHOCK (ANR-17-CE31-0006-01) and FP7 HELCATS project \url{https://www.helcats-fp7.eu/}. P.H., R. H., G.S., and A.V. acknowledge support from the NASA \textit{PSP} program office. 


\appendix 

\section{A New Model for the Internal Magnetic Field} 

The toroidal current of the FR is simplified to a current loop (ring) of major radius $R$, and both toroidal and poloidal currents are allowed to flow inside a minor radius $a$, just as in \cite{Chen1989ApJ,Chen1996JGR}. The toroidal and poloidal currents with densities $J_t$ and $J_p$ generate, respectively, a toroidal ($B_t$) and a poloidal ($B_p$) field that form magnetic field lines wound around the current loop. These magnetic field lines form the toroidal structure. There are no currents outside the FR ($r>a$), nor toroidal magnetic field ($B_t=0$), such that outside the FR, magnetic field lines are fully poloidal and potential. The following development starts from a given FR shape configuration, with a given FR major radius $R$ and minor radius $a$, and then determines the stable magnetic field structure locally. \\

Starting with the vector potential $\vec{A}$ defined as $\vec{B}=\vec{\nabla}\times \vec{A}$ and assuming the FR is axi-symmetric locally, we only need to define the poloidal flux function $\tilde{A}=R\times A_{\varphi}$ where $\tilde{A}=\tilde{A}(r,\theta)$. According to our coordinate system, the magnetic field components can then be expressed in terms of $\tilde{A}$ according to:
			
			\begin{equation}
			\label{eq:B_components_equation}
				\left( B_r, B_\theta, B_\varphi \right) = \frac{1}{R + r cos\theta}  \left( -\frac{1}{r}\frac{\partial\tilde{A}}{\partial \theta} , b_\varphi (\tilde{A}) ,  \frac{\partial\tilde{A}}{\partial r}   \right)
			\end{equation}

such that equation $\vec{\nabla} \cdot \vec{B} = 0$ is automatically satisfied since the divergence of a curl is always zero.\\
		
The magnetostatic equilibrium equation $\vec{j}\times \vec{B} = \vec{\nabla} p $ gives, with the definition of the current density $\mu_0 \vec{j} = \vec{\nabla} \times \vec{B}$, the Grad-Shafranov equation:
			\begin{widetext}
			\begin{equation}
				\label{eq:Grad_Shafranov_eq}
				\left[ \frac{1}{r}\frac{\partial}{\partial r}\left( r \frac{\partial\tilde{A}}{\partial r} \right) + \frac{1}{r^2} \frac{\partial^2\tilde{A}}{\partial \theta^2} \right] - \frac{1}{R+r cos\theta} \left( cos\theta \frac{\partial\tilde{A}}{\partial r} - \frac{sin\theta}{r} \frac{\partial\tilde{A}}{\partial \theta} \right) = -\mu_0 \left( R +r cos\theta \right)^2 \frac{d P}{d \tilde{A}} - \frac{d}{d \tilde{A}}\left( \frac{b^2_\varphi}{2}\right)
			\end{equation}
			\end{widetext}
			
In order to get an analytical solution to the Grad-Shafranov equation, we express the poloidal flux function as the sum of a zeroth-order and a first-order term:
			
			\begin{equation}
				\tilde{A}(r,\theta) = \tilde{A}_0(r) + \tilde{A}_1(r,\theta) 
			\end{equation}
			
such that the zeroth-order component represents the symmetric part of the solution (corresponding to the case of a cylinder) and the first-order $\tilde{A}_1$ term, a function of $\theta$, contains the asymmetric aspects enforced by the toroidal geometry. Replacing in the Grad-Shafranov equation \ref{eq:Grad_Shafranov_eq} and developing in powers of $a/R<<1$, two equations can be obtained: one for the symmetric field (zeroth-order), and a second one for the asymmetric field (first-order) \citep{Priest2014masu}. \\

We extend the derivations made by \citet{Priest2014masu} by assuming a zeroth-order toroidal field as below:
			\begin{equation}
			\begin{aligned}
			\label{eq:BT0_profile}
				&B_{t0}=3\bar{B_{t0}}\left( \frac{a_2}{a}\right)^2\left[ 1-2\left(\frac{r}{a}\right)^2+\left(\frac{r}{a}\right)^4 \right] \\
				&\forall r<a \ \ \text{and} \  B_{t0}=0 \ \ \forall a\leq r 
		    \end{aligned}
			\end{equation}
where $a_2$ is the cross-section radius at the apex of the CME and $\bar{B_{t0}}$ is the average zeroth-order toroidal field in the cross section. After a lengthy derivation, which assumes a uniform plasma pressure inside the FR, the components of the magnetic field inside the cross section of the FR can be obtained and expressed in toroidal coordinates ($R,r,\theta$):

\begin{widetext}	
\begin{equation}
    \label{Inside_field_equation}
				\left\{
				\begin{aligned}
				B_r(r,\theta)&=\frac{-R}{R+rcos\theta}\frac{\Delta(r)}{r}B_{\theta0}(r)sin\theta \ \ \forall 0<r\leq a \\
				B_r(r=0,\theta)&=-3\sqrt{2}\bar{B_{t0}}\left(\frac{a_2}{a}\right)^2\frac{\Delta(r=0)}{a}sin\theta \\
				B_\theta(r,\theta)&=\frac{R}{R+rcos\theta}\left[B_{\theta0}(r)\left(1-cos\theta\frac{d\Delta}{dr}\right)-cos\theta\Delta(r)\frac{dB_{\theta0}}{dr}\right] \ \ \forall 0\leq r\leq a\\
				B_\varphi(r,\theta)&=\frac{R}{R+rcos\theta}B_{\varphi_0}(r)=\frac{R}{R+rcos\theta}B_{t_0}(r) \ \ \forall 0\leq r\leq a \\
				\end{aligned}
				\right.
\end{equation}
\end{widetext}

where $\Delta (r)$ is the Shafranov shift and is obtained by solving a separate differential equation given in \citet{Priest2014masu}. This equation can be re-expressed in terms of the assumed zeroth-order toroidal field obtained from equation \ref{eq:BT0_profile}, the major and minor radii of the FR derived from the force balance equation \ref{eq:forcebalance}, and an assumed profile for the plasma pressure of the FR that is here made dependent on the output of the kinematic model \ref{eq:forcebalance}.\\

The shifting of the center of flux surfaces given by $\Delta (r)$ is strong in CMEs with aspect ratio ($a/R$) greater than 0.5 and with significant magnetic fields. The asymmetric component of the field develops mostly in highly energetic events exhibiting strong Lorentz forces. The event of interest in this study carries relatively weak magnetic fields, and thus no significant asymmetric component develops during the eruption and propagation of the structure. The conditions under which the Shafranov shift becomes significant and affects the internal topology of CMEs will be the subject of a future paper. \\



\end{document}